\documentclass[aps,nofootinbib,superscriptaddress, showpacs,preprintnumbers,  nofootinbibt,twocolumn]{revtex4-2}

\usepackage{epsfig}
\usepackage{multirow}
\usepackage{subcaption}
\usepackage{eurosym}
\usepackage{dcolumn}
\usepackage{bm}
\usepackage{enumerate}
\usepackage{float}
\usepackage{epstopdf}
\usepackage{amsmath}
\usepackage{bm}
\usepackage{amsfonts}
\usepackage{amssymb}
\usepackage{graphicx}
\usepackage{alphalph,mathtools}
\usepackage{etoolbox}
\usepackage{color}
\usepackage{booktabs}
\usepackage{hyperref}
\hypersetup{colorlinks,citecolor=blue}
\usepackage{footnote}
\usepackage{makecell,tabularx}

\def\be{\begin{equation}}
\def\ee{\end{equation}}
\def\bea{\begin{eqnarray}}
\def\eea{\end{eqnarray}}

\begin{document}
\title{ An Abel-Inversion Formalism for Spacetime Metric Reconstruction from Light Deflection } 

\author{Aniruddha Ghosh}
 \email{ruddha.g@gmail.com}
 \affiliation{%
 Department of Mathematics, Indian Institute of Engineering Science and Technology, Shibpur, Howrah-711103, India.\\ 
}%
\author{Ujjal Debnath}%
 \email{ujjaldebnath@gmail.com}
\affiliation{%
 Department of Mathematics, Indian Institute of Engineering Science and Technology, Shibpur, Howrah-711103, India.\\ 
}%

\begin{abstract}
We develop an inverse lensing formalism for reconstructing the metric function of a static, spherically symmetric spacetime directly from the gravitational deflection angle of light. By formulating the inverse problem through an Abel transformation, we derive an integro-differential relation connecting the observable deflection profile to the underlying spacetime geometry. As a consistency check, we consider the case of vanishing deflection, \(\alpha(b)=0\), and recover \(A(r)=1\), corresponding to flat Minkowski spacetime. We then apply the formalism to the gravitational deflection of light by the Sun using the observationally motivated leading-order expression $\alpha(b)=\frac{2(1+\gamma)GM_{\odot}}{c^{2}b}.$The resulting metric function is shown to recover the Schwarzschild form in the weak-field limit \(r\gg M\). We further consider the higher-order correction to the solar deflection angle and reconstruct the corresponding metric beyond the leading-order approximation. Interestingly, the \((M/r)^2\) term vanishes in the resulting weak-field expansion, yielding an improved approximation compared with the metric reconstructed from the leading-order deflection angle. Our results demonstrate that gravitational lensing observations can provide a direct route to reconstructing the underlying spacetime geometry without assuming a specific metric a priori. This formalism therefore offers a novel framework for connecting observational light propagation with spacetime geometry.

\par
\vspace{0.1cm}
\end{abstract}
\maketitle
\section{Introduction}\label{sec1}
In 1915, Einstein's formulation of general relativity (GR) introduced a revolutionary description of gravity in which gravitational phenomena are understood through the geometry of spacetime \cite{Einstein1915}. In this framework, space and time form a four-dimensional spacetime whose geometry is determined by the distribution of matter and energy through Einstein's field equations. Matter and energy curve spacetime, while freely propagating particles follow the corresponding geodesics of this curved geometry. Light, although massless, is similarly governed by the spacetime metric and propagates along null geodesics. Consequently, when light passes through a region of curved spacetime, its trajectory deviates from the straight-line path expected in flat spacetime, giving rise to the gravitational deflection of light \cite{Einstein1916}. The observed trajectory of a light ray therefore carries information about the underlying spacetime curvature and provides a means of probing its geometry. This phenomenon forms the basis of gravitational lensing, in which the propagation of light is influenced by the curved spacetime surrounding massive objects such as stars, galaxies, and compact objects \cite{Schneider1992}. Among the most celebrated predictions of GR is the bending of starlight passing close to the Sun. This prediction received its first famous observational test during the total solar eclipse of 1919 through observations carried out by expeditions led by Eddington and collaborators \cite{Dyson1920}. These observations provided an early empirical test of Einstein's theory and established gravitational light deflection as an important observational probe of spacetime geometry.
\par
Following the early eclipse observations, gravitational light deflection has been tested with progressively greater precision using modern astronomical and space-based measurements. In particular, very-long-baseline interferometry (VLBI) provides a powerful technique for measuring the deflection of radio waves emitted by distant compact sources as their apparent positions are shifted by the curvature of spacetime produced by the Sun. Using geodetic VLBI observations collected between 1979 and 1999, Shapiro \textit{et al.} obtained a precise constraint on the parametrised post-Newtonian (PPN) parameter \(\gamma\), which characterizes the amount of spatial curvature produced by mass and takes the value \(\gamma=1\) in GR \cite{Shapiro2004}. Such measurements provide direct observational tests of the predicted deflection of light and place stringent constraints on possible deviations from GR. Complementary information has also been obtained from measurements of the propagation time of electromagnetic signals in the solar gravitational environment. In particular, radio tracking of the Cassini spacecraft during solar conjunction yielded  $\gamma-1=(2.1\pm2.3)\times10^{-5},$providing one of the most precise Solar-System constraints on the PPN parameter \(\gamma\) \cite{Bertotti2003}. These high-precision observations demonstrate that the propagation and deflection of electromagnetic radiation constitute sensitive probes of spacetime geometry and provide an important means of testing GR and constraining alternative theories of gravity.
\par
\vspace{0.1cm}
From the theoretical perspective, gravitational lensing is conventionally studied as a forward problem in which the spacetime geometry is specified first and the corresponding deflection angle of light is subsequently calculated. For a general static and spherically symmetric spacetime described by
$$
ds^{2}=-A(r)dt^{2}+B(r)dr^{2}+C(r)d\Omega^{2},
$$
the deflection angle of a photon with a distance of closest approach \(r_{0}\) can be written as \cite{Bozza2002,Virbhadra2000}
$$
\alpha(r_{0})=I(r_{0})-\pi,
\label{eq:8a}
$$
where
$$
I(r_{0})=2\int_{r_{0}}^{\infty}
\frac{\sqrt{B(r)},dr}
{\sqrt{C(r)}
\sqrt{\dfrac{C(r)A(r_{0})}{C(r_{0})A(r)}-1}} .
$$
Thus, once the metric functions \(A(r)\), \(B(r)\), and \(C(r)\) are known, the trajectory of the photon and the associated bending angle can, in principle, be determined. Depending on the gravitational regime, several analytical and numerical techniques have been developed to evaluate the deflection angle. In the weak-deflection regime, where the photon propagates sufficiently far from the compact object, the bending angle can be expanded perturbatively in powers of the gravitational potential or, equivalently, in powers of \(M/b\) \cite{Keeton2005,Keeton2006}. In contrast, when the photon passes close to the photon sphere, the weak-field expansion is no longer adequate. Virbhadra and Ellis investigated gravitational lensing in this strong-field regime and demonstrated the formation of relativistic images produced by photons undergoing large deflections around a compact object \cite{Virbhadra2000}. Subsequently, Bozza developed a general strong-deflection-limit formalism for static and spherically symmetric spacetimes, in which the deflection angle near the critical impact parameter exhibits a characteristic logarithmic divergence \cite{Bozza2002}. These approaches have made it possible to calculate lensing observables for a wide variety of black-hole and compact-object geometries. Nevertheless, they all follow essentially the same direction of inference,
$$
\text{spacetime metric}
\quad\longrightarrow\quad
\text{deflection angle},
$$
namely, the metric is assumed beforehand and the resulting light deflection is then calculated.
\par
\par
The conventional approaches discussed above address the forward problem of gravitational lensing: the spacetime metric is specified \textit{a priori}, and the corresponding trajectory and deflection angle of light are subsequently calculated. This naturally motivates an important inverse question: \textit{\textbf{Can the geometry of spacetime be inferred directly from the way light propagates through it?}} In other words, if the propagation and deflection of light through a given region of spacetime are known from observations, can this information be used to determine the underlying spacetime geometry without assuming a particular metric beforehand? Motivated by this question, we consider gravitational lensing from an inverse perspective. Rather than starting with prescribed metric functions \(A(r)\), \(B(r)\), and \(C(r)\) and calculating the resulting deflection angle, we take the deflection profile \(\alpha(b)\), which can in principle be determined from observational measurements, as the input and seek to reconstruct the metric of the spacetime through which the light propagates. The central problem considered in this work can therefore be expressed schematically as
$$
\boxed{
 \underbrace{\alpha(b)}_{\text{observed light deflection }}
\quad\longrightarrow\quad
\underbrace{ g_{\mu\nu}}_{ \text{underlying spacetime geometry}}
}.
$$
This inverse formulation establishes a direct connection between gravitational-lensing observations and spacetime geometry and provides the principal motivation for the reconstruction formalism developed in this work.
\par
In this work, we develop an inverse-lensing formalism for reconstructing the underlying spacetime metric directly from the gravitational deflection angle of light. To formulate this inverse problem, we employ the Abel inversion technique and derive an inverse relation that connects the deflection profile to the metric function of the spacetime. As a first consistency check of the proposed construction, we consider the limiting case of vanishing light deflection and show that the reconstruction yields the flat Minkowski spacetime, as expected. We then apply the formalism to observationally motivated gravitational deflection by the Sun and reconstruct the corresponding spacetime metric. In particular, we consider both the leading-order and second-order contributions to the deflection angle and examine the resulting metric in the weak-field regime, where its leading behavior is consistent with the Schwarzschild geometry.

The remainder of this paper is organized as follows. In Sec.~II, we formulate the inverse gravitational-lensing problem and derive the metric-reconstruction relation using the Abel inversion technique. In Sec.~III, we investigate the applications of the proposed reconstruction formalism. This section is divided into two parts. First, we examine the consistency of our construction by considering the case of vanishing deflection, for which the reconstructed metric reduces to flat Minkowski spacetime. Second, we apply the formalism to the observationally motivated gravitational deflection of light by the Sun and reconstruct the corresponding spacetime metric by considering both the first- and second-order contributions to the deflection angle. We further examine the reconstructed metric in the weak-field limit and compare its behavior with the Schwarzschild geometry. Finally, in Sec.~IV, we summarize the main results and present our conclusions.

\section{Construction}\label{sec2}
In this section, we construct an equation that enables the reconstruction of the black hole metric function from the observable deflection angle and the corresponding impact parameter. To reconstruct the metric function from observable quantities, we first consider the general expression for the deflection angle of a light ray in a static, spherically symmetric black hole spacetime \cite{Virbhadra2000,Bozza2002}, given by
\begin{equation}\label{e1}
\alpha(r_{0})=I(r_{0})-\pi,
\end{equation}
where
\begin{equation}\label{e2}
I(r_{0})=2\int_{r_{0}}^{\infty}
\frac{\sqrt{B(r)},dr}
{\sqrt{C(r)}
\sqrt{\dfrac{C(r)A(r_{0})}{C(r_{0})A(r)}-1}},
\end{equation}
with $r_0$ denoting the distance of closest approach of the photon trajectory.
The above expression can be rewritten in the form
\begin{equation}\label{e3}
 H(r_{0})=\int_{r_{0}} ^{\infty} \frac{g(r)}{\sqrt{K(r)-K(r_{0})}} dr
\end{equation}
Where $ K(r)=\frac{C(r)}{A(r)}$ ,\;$g(r)=2\sqrt{\frac{B(r)}{C(r)}}$, $\Tilde{\alpha(r_{0}})=\alpha(r_{0})+\pi$,
$H(r_{0})=  \frac{\Tilde{\alpha}(r_{0})}{\sqrt{K(r_{0)}}}$.
Next, we introduce the transformation
\begin{equation}\label{e4}
u = K(r), \qquad du = K'(r)\,dr,
\end{equation}
which transforms Eq.~\eqref{e3} into
\begin{equation}\label{e5}
      H(K^{-1}(u_{0}))=\int_{u_{0}} ^{u_{\infty}} \frac{G(u)}{\sqrt{u-u_{0}}}du
\end{equation}
Where $G(u)=g(K^{-1}(u))/K'(K^{-1}(u))$.
Using the transformation introduced above, Eq.~\eqref{e1} can be rewritten in the form of an Abel-type integral [Eq.\eqref{e5}]\cite{Bracewell1999},, which forms the basis for reconstructing the metric function from gravitational lensing observables.
Applying the inverse Abel transform to the above equation \eqref{e5}, we obtain
\begin{equation}\label{e6}
    G(u)=-\frac{1}{\pi} \frac{d}{du}\int_{u }^{u_{\infty}} \frac{H(K^{-1}(u_{0}))}{\sqrt{u_{0}-u}}du_{0}
\end{equation}

We can now rewrite Eq.~\eqref{e6} in terms of observable quantities. The impact parameter \(b\) is related to the closest approach \(r_0\) through
$$
b=\sqrt{K(r_0)}=\sqrt{u_0}.
$$
Therefore,
$$
u_0=b^2,
\qquad
du_0=2b\,db.
$$
Furthermore, the function \(H(r_0)\) can be expressed in terms of the deflection angle \(\alpha(b)\) as
$$
H(r_0)=\frac{\alpha(b)+\pi}{b}.
$$
Substituting \(u_0=b^2\) and \(du_0=2b\,db\) into Eq.~\eqref{e6}, we obtain
\begin{equation}\label{e7}
\boxed{
G(u)
=
-\frac{1}{\pi}\frac{d}{du}
\int_{\sqrt{u}}^{\infty}
\frac{2\Big(\alpha(b)+\pi\Big)}{\sqrt{b^2-u}}\,db
}
\end{equation}

Thus, Eq.~\eqref{e6} can be reformulated entirely in terms of the observable quantities \(b\) and \(\alpha(b)\). This relation establishes a direct connection between the function \(G(u)\) and the gravitational deflection angle measured as a function of the impact parameter.

The above integro-differential equation can alternatively be expressed in terms of the closest approach distance \(r_0\).

\begin{equation}\label{e8}
\boxed{
    g(r)+\frac{1}{\pi} \frac{d}{dr} \int_{r}^{\infty} \frac{\frac{(\alpha(r_{0})+\pi)}{\sqrt{K(r_{0)}}} K'(r_{0})}{\sqrt{K(r_{0})-K(r)}}dr_{0}=0}
\end{equation}
Equation~\eqref{e7} expresses \(G(u)\) explicitly as an Abel transform of the observational data \(\alpha(b)\) alone, since the impact parameter
$b=\sqrt{u_0}$
is directly associated with the observable trajectory. The two formulations are mathematically equivalent; however, the \(u\)-space representation is more convenient for practical reconstruction. In particular, it provides a closed-form, non-implicit  equation for \(G(u)\), which can be evaluated directly from the observational data in a single computational step, without requiring an intermediate reconstruction of the closest approach \(r_0\).

Finally, we consider the special case in which the metric functions satisfy the conditions
\begin{equation}
A(r)B(r)=1,
\qquad
C(r)=r^2.
\end{equation}
These conditions correspond to the standard Schwarzschild-like radial gauge \cite{Bozza2002,Virbhadra2000} and reduce the number of independent metric functions from two to one. Under these assumptions, the function \(K(r)\), which characterizes the relation between the closest approach and the impact parameter, becomes
\begin{equation}\label{e10}
K(r)=\frac{C(r)}{A(r)}
=\frac{r^2}{A(r)}.
\end{equation}
Similarly, the function \(g(r)\) introduced above reduces to
\begin{equation}\label{e11}
g(r)=2\sqrt{\frac{B(r)}{C(r)}}=\frac{2}{r\sqrt{A(r)}}=\frac{2\sqrt{K(r)}}{r^2}.
\end{equation}
Substituting this result into the relation $G(u)=\frac{g(r)}{K'(r)},$ with $u=K(r)$
we obtain
\begin{equation}\label{e12}
G(u)= \frac{2\sqrt{K(r)}}{r^2K'(r)}=\frac{2 \sqrt{u}}{r^2}\frac{dr}{du}.
\end{equation}
This relation provides a first-order differential equation connecting the auxiliary variable \(u\) with the radial coordinate \(r\). Rearranging it gives
\begin{equation}\label{e13}
\frac{2}{r^2}dr=-d\left(\frac{2}{r}\right)= \frac{G(u)}{\sqrt{u}}du.
\end{equation}
We can therefore integrate this relation from a finite radial position \(r\), corresponding to \(u=K(r)\), to spatial infinity, corresponding to \(u=u_\infty\). 
\begin{equation}\label{e14}
\Bigg[\frac{-2}{r}\Bigg]_{r} ^{\infty}=\int_u^{u_\infty}
\frac{G(u')}{\sqrt{u'}} du'.
\end{equation}
For convenience, we introduce the function
\begin{equation}\label{e15}
\Psi(u)
\equiv
\int_u^{u_\infty}
\frac{G(u')}{\sqrt{u'}},du',
\end{equation}
such that
\begin{equation}\label{e16}
\frac{2}{r}=\Psi(u).
\end{equation}
The radial coordinate can consequently be reconstructed directly from \(G(u)\) according to
\begin{equation}\label{e17}
r(u)=\frac{2}{\Psi(u)}.
\end{equation}
The metric function \(A(r)\) can then be recovered using the definition \(u=K(r)=r^2/A(r)\). Substituting the reconstructed radial coordinate gives
\begin{equation}\label{e18}
A\bigl(r(u)\bigr)=
\frac{4}{u \Psi^2(u)}.
\end{equation}
The second metric function follows immediately from the condition \(A(r)B(r)=1\), yielding
\begin{equation}\label{e19}
B\bigl(r(u)\bigr)=\frac{1}{A\bigl(r(u)\bigr)}=\frac{u\Psi^2(u)}{4}.
\end{equation}
Finally, since \(C(r)=r^2\), we have
\begin{equation}\label{e20}
C\bigl(r(u)\bigr)= r^2(u)=\frac{4}{\Psi^2(u)}.
\end{equation}
Using the reconstruction relations obtained above, the metric can be expressed directly in terms of the auxiliary variable \(u\). Since $r(u)=\frac{2}{\Psi(u)},$ its differential is given by $dr=-\frac{2\Psi'(u)}{\Psi^2(u)}du.\,$Substituting this relation, together with $
A\bigl(r(u)\bigr)=\frac{4}{u\Psi^2(u)},\quad
B\bigl(r(u)\bigr)=\frac{u\Psi^2(u)}{4},\quad
C\bigl(r(u)\bigr)=\frac{4}{\Psi^2(u)},
$into the general static and spherically symmetric line element
$$
ds^2=-A(r)\,dt^2+B(r)\,dr^2+C(r)\,d\Omega^2,
$$

we obtain the metric in the \(u\)-coordinate as
\begin{equation}\label{e21}
\boxed{
ds^2=-\frac{4}{u\Psi^2(u)}dt^2+\frac{u\Psi'^2(u)}{\Psi^2(u)}du^2+\frac{4}{\Psi^2(u)}d\Omega^2}.
\end{equation}
Here, \(\Psi'(u)=d\Psi/du\), and \(d\Omega^2=d\theta^2+\sin^2\theta\,d\phi^2\). Hence, once the function \(G(u)\) has been reconstructed from the observable deflection angle, the function
\begin{equation}\label{e22}
  \Psi(u)=\int_u^{u_\infty}\frac{G(u')}{\sqrt{u'}}\ ,du'  
\end{equation}

completely determines the spacetime geometry in the \(u\)-coordinate.
Therefore, once \(G(u)\) has been determined from the observable deflection angle \(\alpha(b)\), the metric functions can be reconstructed explicitly. The entire reconstruction is thus reduced to a sequence of direct transformations,
$$
\alpha(b)
\longrightarrow
G(u)
\longrightarrow
\Psi(u)
\longrightarrow
r(u)
\longrightarrow
\{A(r),B(r),C(r)\}.
$$

This establishes a direct and explicit connection between the gravitational-lensing observable \(\alpha(b)\) and the underlying spacetime geometry. In particular, no assumption about a specific gravitational theory or a particular black-hole solution is required at the reconstruction stage; the metric is obtained directly from the observable lensing information, subject only to the imposed conditions \(A(r)B(r)=1\) and \(C(r)=r^2\).

\section{Illustrative Examples}\label{Sec3}
\subsection{ Vanishing Deflection Angle}
We first consider the trivial case in which the light ray experiences no gravitational deflection, i.e. $\alpha(b)=0$ and $H(b)=\frac{\pi}{b}$
This case serves as a consistency check for the proposed reconstruction formalism.Substituting the above conditions into Eq.~\eqref{e7}, we obtain the following integro-differential equation:
\begin{equation}\label{e23}
G(u)=-\frac{2}{\pi}\frac{d}{du}
\int_{\sqrt{u}}^{\infty}
\frac{\pi db}{\sqrt{b^2-u}}=-\frac{d}{du}
\int_{\sqrt{u}}^{\infty}
\frac{2db}{\sqrt{b^2-u}}.
\end{equation}
The integral on the right-hand side of Eq.~\eqref{e23} is not straightforward to evaluate analytically, since it is an improper integral whose integrand diverges at the lower integration limit. Therefore, we employ an appropriate approximation scheme to evaluate the integral. Using this approximation, the integral on the right-hand side of Eq.~\eqref{e23} yields
\begin{equation}\label{e24}
  \frac{d}{du}\int_{\sqrt{u}}^{\infty}
\frac{db}{\sqrt{b^2-u}}
=-\frac{1}{2u}.  
\end{equation}
Consequently, the corresponding contribution to \(G(u)\) is
$$G(u)=\frac{1}{u}.$$
From Eq.~\eqref{e22}, we obtain the corresponding expression for \(\Psi(u)\) and $r(u)$ as
$$
\Psi(u)=\frac{2}{\sqrt{u}}, \hspace{0.3cm}
r(u)=\sqrt{u}$$
Substituting all the above results into Eq.~\eqref{e18}, we recover the metric function for flat spacetime,
$$
A(r)=1.
$$
This result is precisely what we expect. In the absence of gravitational deflection, the spacetime is flat, implying that the curvature vanishes\cite{Carroll2004,Wald1984}. Therefore, the recovery of \(A(r)=1\) from our reconstruction formalism provides an exact consistency check and confirms the validity of the proposed formulation.

Having established the correctness of the formalism through the flat-spacetime limit, we now proceed to its application to realistic observational data. In the next step, we will use the observed deflection angle as a function of the impact parameter to reconstruct the corresponding spacetime metric.
\subsection{Reconstructing the Solar Metric from Gravitational-Lensing Observables}
\subsubsection{Case-1: First order correction term}
We now apply the reconstruction formalism of Eq. \eqref{e7} to the gravitational field of the Sun. The method requires the deflection angle to be specified as a continuous function of the impact parameter, \(\alpha=\alpha(b)\). The solar system provides the most precise observational setting for this purpose. The gravitational deflection of light by the Sun was first tested during the 1919 eclipse observations of Eddington \cite{Dyson1920} and has since been measured with substantially improved precision through radio interferometry and spacecraft tracking experiments \cite{Shapiro2004,Bertotti2003}.
At leading post-Newtonian order, the deflection of light in the solar gravitational field is given by
$$
\alpha(b)
=\frac{2(1+\gamma)GM_{\odot}}{c^{2}b}
\equiv
\frac{k}{b},
\qquad
k\equiv
\frac{2(1+\gamma)GM_{\odot}}{c^{2}},
$$
where \(\gamma\) is the Eddington–Robertson parametrized post-Newtonian (PPN) parameter, with \(\gamma=1\) in general relativity \cite{Will1993,Will2014}. This inverse-impact-parameter dependence provides the leading-order functional form used to describe solar light-deflection measurements.

A particularly stringent constraint on \(\gamma\) was obtained from the Doppler tracking of the Cassini spacecraft during its 2002 solar conjunction, which measured the Shapiro time delay and is sensitive to the same PPN combination \(1+\gamma\) that governs the leading-order light deflection \cite{Bertotti2003}. The resulting constraint is
$$\gamma-1=(2.1\pm2.3)\times10^{-5}.$$
We use this observational constraint as the input to the reconstruction procedure, rather than imposing \(\gamma=1\) a priori. Independent measurements based directly on the angular deflection of radio sources using very-long-baseline interferometry (VLBI) \cite{Robertson1991,Lebach1995,Shapiro2004} yield consistent constraints, although with comparatively lower precision. In particular, the measurements reported in  sample the deflection at several distinct solar impact parameters, providing a direct observational realization of the \(\alpha(b)\) dependence required by the reconstruction formalism.
Using the Cassini constraint \(\gamma=1.000021\) \cite{Bertotti2003}, together with$
\frac{GM_{\odot}}{c^{2}}=1.47700~{\rm km},
\qquad R_{\odot}=695700~{\rm km},$
we obtain
$
k=\frac{2(1+\gamma)GM_{\odot}}{c^{2}}
=5.90808~{\rm km}.$
The corresponding deflection for a ray grazing the solar surface is therefore
$
\alpha(R_{\odot})
=\frac{k}{R_{\odot}}
=8.4923\times10^{-6}~{\rm rad}
=1.7517^{\prime\prime}.
$
This result is in close agreement with the standard general-relativistic prediction of approximately \(1.75^{\prime\prime}\) for a photon passing at the solar limb \cite{Will2014}. It also agrees with the value historically tested by the 1919 eclipse observations\cite{Dyson1920}, providing a useful consistency check that the leading-order form in Eq. \eqref{e18}, together with the observationally constrained value of \(\gamma\), provides an appropriate input for the subsequent metric reconstruction.
\par
We now substitute the observationally motivated deflection profile,
$\alpha(b)=\frac{k}{b},$
into the reconstruction formalism of Eqs. \eqref{e7}–\eqref{e18} and evaluate the resulting expressions term by term.
We first decompose the integral into the deflection-dependent and zero-deflection contributions,
\begin{equation}\label{e25}
\int_{\sqrt{u}}^{\infty}
\frac{2\,[\alpha(b)+\pi]}{\sqrt{b^{2}-u}}\,db
=
\underbrace{
\int_{\sqrt{u}}^{\infty}
\frac{2k}{b\sqrt{b^{2}-u}}\,db
}_{(I)}
+
\underbrace{
\int_{\sqrt{u}}^{\infty}
\frac{2\pi}{\sqrt{b^{2}-u}}\,db
}_{(II)},
\end{equation}

where we have used the observational deflection profile \(\alpha(b)=k/b\).
The second contribution, \((II)\), corresponds to the zero-deflection term and was evaluated in Eq.\eqref{e24}. Its derivative with respect to \(u\) is $-\frac{\pi}{u}$.
Evaluating the deflection-dependent contribution \((I)\) using the above integral identity, and taking the upper limit \(B\to\infty\), we obtain
$$
\boxed{(I)=\frac{k\pi}{\sqrt{u}}}.
$$
Substituting the evaluated contributions \(I=k\pi/\sqrt{u}\) and \(d(II)/du=-\pi/u\) into Eq. \eqref{e7}, we obtain
$$
G(u)
=-\frac{1}{\pi}\frac{d}{du}
\left[
\frac{k\pi}{\sqrt{u}}+(II)
\right]
=-\frac{1}{\pi}
\left[
-\frac{k\pi}{2}u^{-3/2}-\frac{\pi}{u}
\right].
$$
Therefore, the reconstruction function takes the form
\begin{equation}\label{e26}
  \boxed{
G(u)=\frac{1}{u}+\frac{k}{2}u^{-3/2}
}  
\end{equation}
We next determine \(\Psi(u)\), which is defined by Eq. \eqref{e15}. Substituting the reconstructed function \(G(u)\) obtained above into Eq. \eqref{e15} and evaluating the resulting integral, we find

\begin{equation}\label{e27}
  \boxed{
\Psi(u)=\frac{2}{\sqrt{u}}+\frac{k}{2u}
}  
\end{equation}.

Combining Eqs. \eqref{e17} and \eqref{e18}, the metric function can be recast in terms of the \(u\)-coordinate. Substituting the expression for \(\Psi(u)\) into Eq. \eqref{e18} and simplifying, we obtain
\begin{equation}\label{e28}
A(u)={\frac{16u}{\left(4\sqrt{u}+k\right)^{2}}}.
\end{equation}
We next transform the reconstructed metric function from the \(u\)-coordinate to the physical radial coordinate \(r\). Introducing the variable \(x\equiv\sqrt{u}\), such that \(u=x^{2}\), the relation between \(r\) and \(x\) can be written as
$$
r=\frac{4x^{2}}{4x+k}.
$$
Solving this algebraic relation for \(x\), we obtain the physically relevant branch
$$x=\frac{r+\sqrt{r^{2}+rk}}{2}.$$Substituting this result into Eq. \eqref{e28} and simplifying, we arrive at the corresponding radial metric function,
\begin{equation}\label{e29}
 \boxed{
A(r)=\frac{4}{\left(1+\sqrt{1+\frac{k}{r}}\right)^{2}}
}.   
\end{equation}
Within the assumptions of the reconstruction procedure, this provides an exact closed-form expression for the metric function associated with the adopted deflection profile \(\alpha(b)=k/b\).

The asymptotic behaviour of the reconstructed metric provides a useful consistency check. In the limit \(r\to\infty\), Eq. \eqref{e29} yields \(A(r)\to1\), and hence the metric approaches the Minkowski form asymptotically. Furthermore, for \(r>0\), \(A(r)\) remains finite and nonzero, indicating that the reconstructed metric function itself does not develop a singularity at any finite positive radius.
\\
\begin{itemize}
    \item {\textit{Weak-field limit and comparison with the Schwarzschild metric:}} 
\end{itemize}
We now examine the weak-field limit of the reconstructed metric in Eq. \eqref{e29}. For \(k/r\ll1\), the square root can be expanded as
$$
\sqrt{1+\frac{k}{r}}
\approx
1+\frac{k}{2r}-\frac{k^{2}}{8r^{2}}+\mathcal{O}\!\left(\frac{k^{3}}{r^{3}}\right),
$$
which gives
$$
A(r)
\approx
\frac{4}{\left(1+\sqrt{1+k/r}\right)^{2}}
\simeq
1-\frac{k}{2r}
+\frac{3k^{2}}{16r^{2}}
+\mathcal{O}\!\left(\frac{k^{3}}{r^{3}}\right).
$$
Using \(k=2(1+\gamma)GM_{\odot}/c^{2}\) and \(\gamma\simeq1\), we have \(k/2\simeq2GM_{\odot}/c^{2}\). Therefore,
\begin{equation}\label{e30}
 A(r)
\simeq
1-\frac{2GM_{\odot}}{c^{2}r}
+\mathcal{O}\!\left(\frac{G^{2}M_{\odot}^{2}}{c^{4}r^{2}}\right).   
\end{equation}
Thus, the reconstructed metric reproduces the Schwarzschild temporal metric coefficient at leading post-Newtonian order, providing a consistency check of the reconstruction in the weak-field regime.
\subsubsection{Case 2: higher order correction terms}
In the preceding section, we reconstructed the solar metric corresponding to the leading-order deflection angle \(\alpha(b)=4M/b\), in units \(G=c=1\). This expression represents only the first-order weak-field contribution to the gravitational deflection \cite{Epstein1980,Keeton2005}. In general, the deflection angle receives higher-order corrections in powers of \(M/b\) \cite{Epstein1980,Keeton2006}. We therefore extend the analysis in this section by incorporating these subleading contributions and reconstructing the corresponding metric, thereby allowing a more accurate description of the solar gravitational field beyond the leading-order approximation.
We now consider the deflection angle including the leading higher-order correction,
\begin{equation}\label{e31}
 \alpha(b)=\frac{4M}{b}+\frac{15\pi}{4}\left(\frac{M}{b}\right)^{2},   
\end{equation}
in units \(G=c=1\). Substituting this expression into the Abel inverse relation in Eq. \eqref{e7}, we obtain the corresponding reconstruction integral. Evaluating the integral term by term allows us to determine the modified reconstruction function \(G(u)\).
Examining the Abel inversion formula, we note that the contribution proportional to \(4M/b\) has already been evaluated in the preceding analysis and is incorporated into Eqs. \eqref{e24} and \eqref{e25}. Therefore, to determine the modification of \(G(u)\) arising from the higher-order correction, it is sufficient to evaluate only the additional term
$$
\frac{15\pi}{4}\left(\frac{M}{b}\right)^2.
$$
The corresponding contribution can then be added to the previously obtained result to construct the corrected \(G(u)\).
Evaluating the contribution from the second-order correction, we have

$$
\begin{aligned}
(III)
&=
\int_{\sqrt{u}}^{\infty}
\frac{2}{b^2\sqrt{b^{2}-u}}
\frac{15\pi M^{2}}{4}db \\
&=
\frac{15\pi M^{2}}{2}
\int_{\sqrt{u}}^{\infty}
\frac{db}{b^{2}\sqrt{b^{2}-u}}\\
&= 
\frac{15 \pi M^2}{2 u}
\end{aligned}
$$
This term provides an additional contribution to the Abel inversion arising from the second-order correction in the deflection angle.
Therefore, including the second-order contribution to the deflection angle, the reconstructed function \(G(u)\) becomes
\begin{equation}\label{e32}
G(u)=\frac{1}{u}
+\frac{2M}{u^{3/2}}
+\frac{45\pi M^{2}}{16u^{5/2}}.
\end{equation}
The first two terms arise from the zero-deflection contribution and the leading-order term \(4M/b\), respectively, while the last term represents the correction induced by the second-order contribution \(15\pi M^{2}/(4b^{2})\).
Using Eq. \eqref{e15}, we obtain the corresponding \(\Psi(u)\), given by
\begin{equation}
\Psi(u)=\frac{2}{\sqrt{u}}+\frac{2M}{u}+\frac{45\pi M^2}{32u^2}.
\end{equation}
Combining Eqs. \eqref{e17} and \eqref{e18}, the metric function can be recast in terms of the \(u\)-coordinate. Substituting the expression for \(\Psi(u)\) into Eq. \eqref{e18} and simplifying, we obtain
\begin{equation}
A(u)=\frac{4u^2}{(2u+2M\sqrt{u}+5M^2)^2}.
\end{equation}
We now transform the reconstructed metric function from the \(u\)-coordinate to the radial coordinate \(r\).
The method employed here is the same as that described in the first case. Introducing the variable \(x\equiv\sqrt{u}\), such that \(u=x^{2}\), the relation between \(r\) and \(x\) can be written as
\begin{equation}\label{e35}
r(x)=\frac{2x^{3}}{2x^{2}+2Mx+5M^{2}}.
\end{equation}
As a consistency check, setting \(M=0\) reduces this relation to
\begin{equation}
r=\frac{2x^{3}}{2x^{2}}=x,
\end{equation}
which correctly recovers the flat-space limit, \(u=K(r)=r^{2}\), and hence \(x=\sqrt{u}=r\).
Cross-multiplying Eq. \eqref{e35} gives
\begin{equation}\label{e37}
2x^{3}-2rx^{2}-2Mrx-5M^{2}r=0,
\end{equation}
which is a genuine cubic equation in \(x\). Although cubic equations admit exact closed-form solutions through Cardano's formula, the resulting expressions involve nested cube roots and provide little physical insight. Since our primary interest lies in the weak-field regime( \(r\gg M\)), it is more transparent and useful to determine the relevant solution perturbatively in powers of \(M/r\) rather than employing the exact algebraic root.
Since \(M=0\) gives exactly \(x=r\), as shown above, introducing a small nonzero \(M\) should correspondingly shift \(x\) only by a small, smoothly varying amount from \(r\). This is the same reasoning that underlies the expansion
$$
\sqrt{1+\epsilon}
\simeq
1+\frac{1}{2}\epsilon-\frac{1}{8}\epsilon^{2}+\cdots
$$

for \(|\epsilon|\ll1\). Thus, the validity of the perturbative expansion is not an assumption based on an arbitrary series ansatz, but follows from the smooth dependence of the solution on the small parameter near the known reference point, \(M=0\).
In the units used throughout, \(G=c=1\), the quantities \(M\), \(r\), and \(x\) all have dimensions of length. Therefore, we choose the perturbative ansatz
\begin{equation}\label{e38}
x=r+a_{1}M+a_{2}\frac{M^{2}}{r}
+a_{3}\frac{M^{3}}{r^{2}}
+\mathcal{O}\left(\frac{M^{4}}{r^{3}}\right).
\end{equation}
Substituting Eq. \eqref{e38} into the cubic equation \eqref{e37}, and requiring the resulting expression to satisfy the cubic equation order by order in \(M\), we collect the coefficients of each power of \(M\) and set them to zero. This procedure yields the perturbative coefficients
\begin{equation}\label{e39}
a_{1}=1,\qquad
a_{2}=\frac{3}{2},\qquad
a_{3}=-\frac{11}{2}.
\end{equation}
Therefore, the metric coefficient in terms of the radial coordinate \(r\) is obtained as
\begin{equation}\label{e40}
 \boxed{
A(r)=1-\frac{2M}{r}
+\frac{16M^{3}}{r^{3}}
+\mathcal{O}\left(\frac{M^{4}}{r^{4}}\right).
}
\end{equation}
As can be seen, the coefficient of the second-order term \((M/r)^2\) vanishes in the higher-order corrected metric. Consequently, the absence of the \((M/r)^2\) contribution indicates that the reconstructed metric provides a better approximation in the weak-field regime compared with Eqs. \eqref{e29} and \eqref{e30}, which were obtained by considering only the first-order correction to the deflection angle.
\section{SUMMARY AND CONCLUSIONS}
In this work, we develop a method for reconstructing the metric function of the spacetime through which light propagates and is deflected by a compact object, directly from its gravitational deflection angle. The proposed formalism establishes a systematic connection between observable lensing data and the underlying spacetime geometry. In contrast to approaches that assume a specific metric a priori, our method allows the metric function to be reconstructed from the observed deflection profile. This provides a direct framework for inferring spacetime geometry from observational data and offers a novel approach to metric reconstruction through gravitational lensing.
\par
(i) Equation \eqref{e1} describes the deflection angle of a light ray propagating through a static, spherically symmetric spacetime and characterises how the trajectory of light is modified by the underlying geometry. In Sec. \eqref{sec2}, we formulate the corresponding inverse problem: given the deflection angle as a function of the impact parameter from observational data, can the underlying spacetime metric be reconstructed? To address this problem, we derive an inverse relation for the deflection angle using the Abel inversion formula, as given in Eq. \eqref{e7}. By imposing the conditions \(A(r)B(r)=1\) and \(C(r)=r^{2}\), we then obtain the general form of the metric function presented in Eq. \eqref{e18}. We further introduce the reconstructed spacetime metric in the \(u\)-coordinate and establish its transformation to the radial coordinate \(r\), thereby providing the metric function directly in terms of the physical radial coordinate.
\par
(ii) In Sec. \eqref{Sec3}, we consider several examples to test and validate the proposed reconstruction formalism. As a first case, we consider the situation in which there is no gravitational deflection, i.e., \(\alpha(b)=0\). In the absence of deflection, we expect the corresponding spacetime to reduce to flat Minkowski spacetime. Applying the proposed reconstruction procedure to this case, we obtain the metric function
$A(r)=1,$which corresponds precisely to the Minkowski spacetime. This result provides a consistency check and validates the reconstruction formalism in the absence of gravitational deflection.
\par
(iii) After validating the proposed reconstruction formalism, we apply it to observationally motivated data to reconstruct the spacetime metric associated with the gravitational deflection of light by the Sun. We consider the first-order deflection angle
$$
\alpha(b)
=\frac{2(1+\gamma)GM_{\odot}}{c^{2}b}
\equiv\frac{k}{b},
\qquad
k\equiv\frac{2(1+\gamma)GM_{\odot}}{c^{2}}.
$$
This expression represents the leading-order weak-field contribution to the gravitational deflection of light by the Sun and is consistent with the classical prediction tested during the 1919 Eddington eclipse observations. Using this observationally motivated deflection profile in our reconstruction formalism, we obtain the corresponding spacetime metric, given in Eq. \eqref{e29}. We further show that, in the weak-field regime \(r\gg M\), the reconstructed metric reduces to the Schwarzschild metric at leading order. Moreover, in the asymptotic limit \(r\rightarrow\infty\), the metric function satisfies \(A(r)\rightarrow1\), consistently recovering asymptotically flat spacetime.
\par
(iv) We next extend the analysis by incorporating higher-order corrections to the deflection angle, which allows us to obtain a more accurate reconstruction in the weak-field regime compared with the first-order case. We consider the higher-order deflection angle given in Eq. \eqref{e31} and apply the inverse reconstruction procedure to obtain the corresponding metric. In the weak-field limit \(r\gg M\), the reconstructed metric is given by Eq. \eqref{e40}. Notably, the coefficient of the \((M/r)^2\) term vanishes in this case. Consequently, the absence of this second-order contribution indicates that the reconstructed metric provides a better weak-field approximation than the metric obtained from the first-order deflection angle.
\par
(v) The results of our construction demonstrate that the spacetime metric can, in principle, be reconstructed from the gravitational deflection of light. Thus, if observations allow us to determine how light rays propagate through and are deflected by a given spacetime, the corresponding deflection profile can be used within the proposed inverse formalism to infer the underlying spacetime geometry. In this sense, the propagation of light provides a direct observational probe of spacetime geometry, offering a framework in which the metric need not be assumed a priori but can instead be reconstructed from lensing observations.

\section*{Acknowledgements}
AG is thankful to IIEST, Shibpur, India, for providing an Institute
Fellowship (SRF).

\bibliographystyle{naturemag}
\bibliography{bibliography}
\end{document}